\documentclass[12pt,preprint]{aastex}
\usepackage[]{natbib}
\usepackage{graphicx}
\usepackage{datetime,color}

\begin{document}

\shorttitle{}
\shortauthors{}

\title{The Classificiation of \textit{Kepler} B star Variables}

\author{Bernard J. McNamara}
\affil{Department of Astronomy, New Mexico State University, P.O. Box 30001, MSC
4500, Las Cruces, NM 88003-8001}
\email{bmcnamar@nmsu.edu}

\author{Jason Jackiewicz}
\affil{Department of Astronomy, New Mexico State University, P.O. Box 30001, MSC
4500, Las Cruces, NM 88003-8001}
\email{jasonj@nmsu.edu}

\author{Jean McKeever}
\affil{Department of Astronomy, New Mexico State University, P.O. Box 30001, MSC
4500, Las Cruces, NM 88003-8001}

\begin{abstract}

The light curves of 252 B-star candidates in the Kepler data base are analyzed in a similar fashion to that done by Balona et al. (2011) to further characterize B star variability, increase the sample of variable B stars for future study, and to identify stars whose power spectra include particularly interesting features such as frequency groupings. Stars are classified as either constant light emitters, $\beta$ Cep stars, slowly pulsating B stars, hybrid pulsators, binaries or stars whose light curves are dominated by rotation (Bin/Rot), hot subdwarfs, or white dwarfs.  One-hundred stars in our sample were found to be either light contants or to be variable at a level of less than 0.02 mmag. We increase the number of candidate B star variables found in the Kepler data base by Balona et al. (2011) in the following fashion: $\beta$ Cep stars from 0 to 10, slowly pulsating B stars from 8 to 54, hybrid pulsators from 7 to 21, and Bin/Rot stars from 23 to 82.  For comparison purposes,  approximately 51 SPBs and 6 hybrids had been known prior to 2007 (De Cat, 2007). The number of $\beta$~Cep stars known prior to 2004 was 93 (Stankov $\&$ Handler 2005).  A secondary result of this study is the identification of an additional 11 pulsating white dwarf candidates, four of which possess frequency groupings. 
\end{abstract}

\section{Introduction}

The primary goal of this investigation is to extend the Kepler B-star study conducted by \citet{balona2010} to a substantially larger number of objects. Their paper examined the light curves of 48 B stars in the NASA Kepler satellite database and found that their frequency spectra differed from those obtained from ground-based observatories. Much smaller amplitude, high-degree modes are now detectable because of Kepler's high precision and continuous observing cadence.  This study further quantifies the nature of B-star variability by uniformly analyzing the power spectra of 252 stars, essentially all B-star candidates with $m_{\rm V}<14.5$ in the Kepler field-of-view.

Our study was motivated by the puzzling manner in which B stars pulsate. Their interior structures are thought to be  nearly identical, so why are their pulsational properties different? The $\beta$ Cephei ($\beta$~Cep) stars are typically multi-periodic radial pulsators \citep{stankov2005}. However, the $\beta$~Cep star, $\beta$~Cru appears to pulsate solely in non-radial modes \citep{decat2004}. The spectral range of the $\beta$~Cep stars also overlaps that of the early type Be stars. Nearly all (86\%) of these latter stars show light and line-profile variations indicative of non-radial pulsation, however, only 40$\%$ of mid type Be stars and 18$\%$ of late-type Be stars appear to pulsate in this fashion \citep{neiner2009}. The Be stars also change their light output on a variety of timescales \citep{porter2003}. The spectral range of the slowly pulsating B stars (SPBs) also overlaps that of the $\beta$~Cep \citep{miglio2007}, but their frequency spectra are different. Their spectral range also overlaps that of the Be stars. These diverse observational behaviors may be due to intrinsic differences such as their rotational velocity distributions, age, metal abundance, magnetic field strengths, or degree of core convective overshooting. They might also be caused by external factors such as the presence of a circumstellar disk, our viewing angle to the star, or a binary companion.

The tools of asteroseismoloy can be used to measure many of the above internal stellar properties \citep{decat2004,aerts2008}. Ground-based programs have attempted to do this, but they generally detect too small a number of frequencies. The pioneering space-based WIRE (Wide-Field InfraRed Explorer) and MOST (Microvariability and Oscillating Stars) satellites were considerably more successful in this regard and detected dozens of frequencies \citep{busazi2002,bruntt2007,matthews2004,walker2005b}.  However, neither of these micro-satellites can observe a significant sample of B stars because they are limited to very bright stars ($m_{\rm V}<6$). 

In contrast to the above programs, Kepler simultaneously monitors the light variations of about 150,000 stars. Although its primary mission is to search for eclipses caused by transiting Earth-like planets, dozens of high quality B-star light curves are also being acquired.  The potential that Kepler holds for the study of B-star variability was amply demonstrated by the investigation of \citet{balona2010}. Among the 48 light curves they examined, fifteen were found to be pulsators. This study expands on that pioneering work by examining a much larger set of Kepler B stars and classifying their variability. Therefore, this study is limited to a discussion of the Kepler B-star frequency spectra. The relation between these spectra and a star's position in the HR diagram will be the subject of a later paper. 

This paper is organized in the following fashion. In Section~\ref{data} the Kepler data set is described. Section~\ref{freq}  provides an overview of the frequency analysis. Section~\ref{disc} provides a discussion of the frequency spectra of our target stars and Section~\ref{summ} consists of a summary of our findings.

\section{Data}
\label{data}

The Kepler satellite was launched on March 6, 2009. It monitors the brightness variations of about 150,000 stars in a 105 $\deg^2$  region in the constellations of  Cygnus and Lyra. In preparation for its launch, \citet{brown2011} obtained multi-band photometry (\textit{griz}, DD051, and \textit{JHK} from 2MASS) for more than 4 million stars in this  field. These colors were used to assign each star an effective temperature $T_{\rm eff}$, surface gravity $\log(g)$, visual absorption $A_{V}$, metallicity, and radius. The Kepler team then selected about 150,000 of these stars whose light curves would be continuously measured. The measurements were obtained in a single filter whose effective wavelength is close to that of the Johnson R filter.  Magnitudes in this bandpass are referred to as Kepler magnitudes ($K_p$) in the Kepler Input Catalogue (KIC). Light curve points were obtained at a cadence of 30 minutes (long cadence) or 1 minute (short cadence). The stars monitored each quarter may change as the Kepler team refines their planet search targets and as new objects are added by participants in the Kepler Guest Observers program. Targets are given a unique identification number ranging from 757076 to 12984227.

\begin{figure}
  \centering
  \includegraphics[width=.8\textwidth,clip]{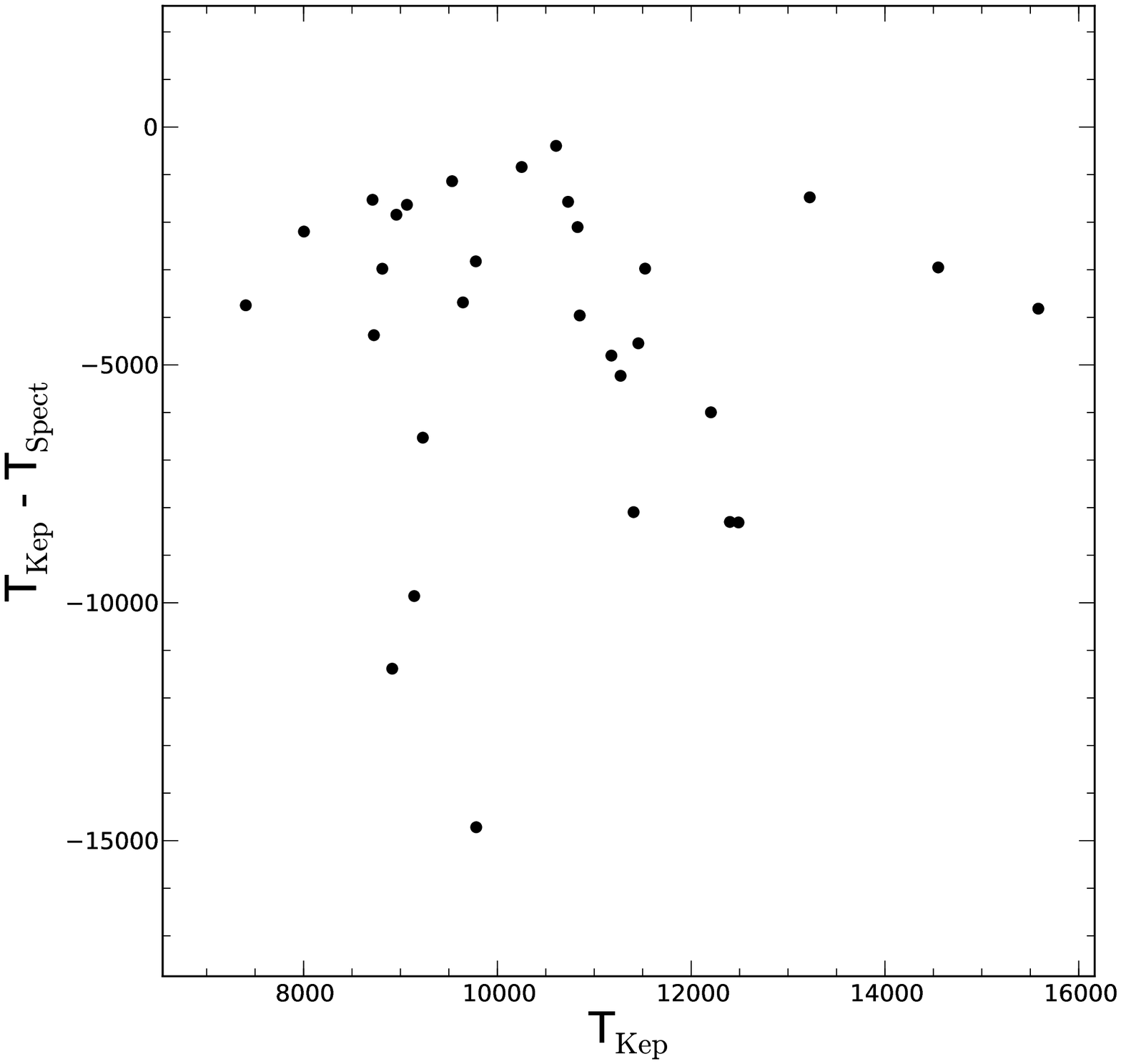}
  \caption{Difference in the Kepler and spectroscopic temperature of a B star versus its Kepler temperature. For hot stars, the effective temperatures in the Kepler database are too low by about 3000~K, but some temperatures can be underestimated by more than 10000~K.}
  \label{fig:temp}
\end{figure}

\begin{figure}
  \centering
  \includegraphics[width=.6\textwidth,clip]{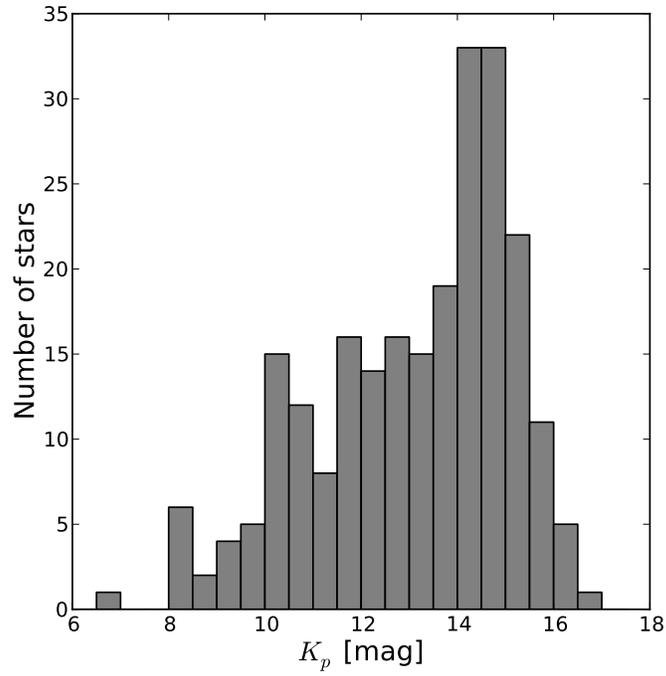}
  \caption{The histogram of our 252 program stars peaks at a Kepler magnitude between 14 and 15. Fainter B-star candidates have a lower signal-to-noise making it more difficult to detect small amplitude variations in their power spectra.}
  \label{fig:maghist}
\end{figure}

This study targets B stars. Program objects were identified by requiring (1) that they have a $T_{\rm eff}>10,000$~K in the KIC, or (2) a position in a KIC color-color plot that is close to that of a B star.  The latter criterion was adopted because the Kepler temperatures of upper main sequence stars are not very precise \citep{lehmann2011} . As a consequence of criterion 2, our initial sample was expected to contain some early-type A stars. In Figure \ref{fig:temp} we compare the effective temperature of B stars whose temperatures were measured spectroscopically in the \citet{balona2010} study to those listed in the KIC. In general the Kepler B-star effective temperatures are too low by about 3000~K, however, in extreme cases these temperatures can differ by a factor of two.  Because the error in $\log(g)$ is relatively large, no restriction was placed on its value when creating our initial list of B-star candidates. The application of these criteria resulted in a sample size of 252 stars. An examination of their power spectra revealed that many (100) candidates were not light variables or possessed a very low level of variability. The KIC numbers of these stars are listed in Table~\ref{nonvar}. If their number is followed by an asterisk, they might be variable at a level of less than about 0.02~mmag.  The properties of the remaining 152 stars in our sample are listed in Table~\ref{bstar}. The first column provides their KIC number. The remaining columns provide the star's Kepler magnitude ($K_p$), Sloan $g$ magnitude and $g-i$ color, its proper motion in right ascension and declination, and a pseudo absolute magnitude, $M_\mu$, that is used later in the text.   A histogram of their $K_p$ magnitudes is shown in Figure~\ref{fig:maghist}.   All of the light curves for our program stars were obtained in the long-cadence mode.

\section{Frequency Analysis}
\label{freq}

The light curves for all 252 program stars were analyzed in a consistent way. All the light-curve measurements  were obtained from the MAST database. We experimented with using the recently released target pixel files and corrections to it, but found that the corrected light curves were more useful for our purposes.

We handled the different epochs by analyzing all available quarters ($n_q$) for each star as well as the entire time series separately, to give $n_q+1$ frequency spectra per star. Outlier points in all light curves were removed, and each time series was interpolated onto an equidistant time grid. Light curves were then put into units of parts-per-million (ppm), using the transformation from the original Kepler flux ($F_{\rm K}$) as $f(t) = 10^6(F_{\rm K}/y - 1)$, where $y$ was either the mean value of the total light curve or a low-order polynomial fit to the light curve, depending on any artifacts present in the data. This fitting did well in removing long-term trends in the light curves that are likely non physical in origin. A similar approach was carried out in \citet{uytterhoeven2011}.

For each case, oversampled power spectra were computed and the frequency of maximum power and its corresponding amplitude were used in a minimization procedure to determine the true frequency, amplitude, and phase (and errors in all three parameters) of a sine function that models the mode. After each iteration, the data were pre-whitened, and the routine continued to determine the frequency set  iteratively until an undesirable signal-to-noise (SNR) ratio of $\le 5$ was reached. This SNR was defined as the amplitude of the pulsation  compared to the median amplitude of the very high frequency regime.


After the frequency analysis of all quarters of all stars, it was necessary to attempt to determine which frequencies were real by eliminating those that could have had instrumental origin. The criterion we used was that a frequency had to be present in each quarter as well as the entire time series spectrum. Additionally, to ensure  the robustness of the frequencies reported in this study, only those having an amplitude within 20\% of the largest amplitude mode were considered. This step certainly under counts the number of frequencies present in many of the stars, but is adequate for determining the type of variability these stars possess. In our view, additional work on the removal of instrumental effects and further high-quality measurements are needed to confirm the presence of the lower-amplitude frequencies.

\section{Discussion}
\label{disc}

\begin{figure}
  \centering
  \includegraphics[width=.7\textwidth,clip]{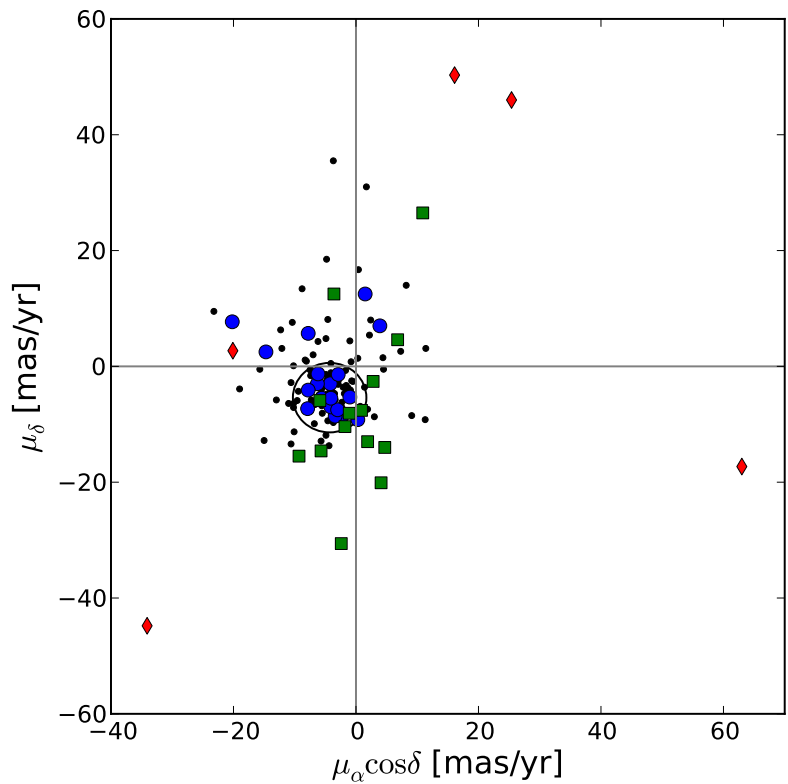}
  \caption{Proper motion in right ascension versus the proper motion in declination of the stars listed in Table~\ref{bstar}, as well as spectroscopically classified non pulsators. Motions were taken from the catalogue of \citet{zacharias2009}. The circle near the center of this figure, at coordinates (-4.3,-5.4)~mas~yr$^{-1}$ and with radius  6~mas~yr$^{-1}$, shows the location of spectroscopically classified B stars denoted as large blue dots.  Green squares denote spectroscopically classified hot subdwarfs  and red diamonds denote white dwarfs. Program stars that have not been spectroscopically classified are denoted by small black circles. }
  \label{fig:vpd}
\end{figure}

The \citet{balona2010} study classified the variable B-stars into 4 groups: SPB stars, SPB/ $\beta$~Cep hybrids, stars with frequency groupings, and stars with variations due to binarity or possibly rotation-modulated magnetic activity. To distinguish between white dwarfs, hot sub dwarfs, and main-sequence B stars, they used proper motions in the UCAC3 catalogue \citep{zacharias2009}. Statistically, stars with large proper motions are expected to be closer to Earth than stars with smaller motions. B stars are rare, therefore, the nearest is expected to be far from us and have a small proper motion.  Proper motions for all but six stars in our sample stars are contained in the UCAC3 catalogue. Figure~\ref{fig:vpd} shows their vector point diagram.  If we only consider stars classified as a B star in  the previous \citep{balona2010,lehmann2011, ostensen2010} studies, their average motion is  (-4.1, -5.1) mas yr$^{-1}$ with a  $1 \sigma$ dispersion of about 1.3  mas yr$^{-1}$.  This centroid is close to the value found by \citet{balona2010} (-4.3, -5.4), and  consistent with that expected of a young star group in the direction of the Kepler field-of-view \citep{kharchenko2003}, so we adopt it for this study. The location of a star in this diagram provides a crude way of distinguishing B stars from other stellar types.  B stars in the upper-right quadrant of the figure have motions consistent with membership in the Gould belt, a partial ring of stars inclined by about 18 degrees to the galactic plane.

\begin{figure}
  \centering
  \includegraphics[width=.7\textwidth,clip]{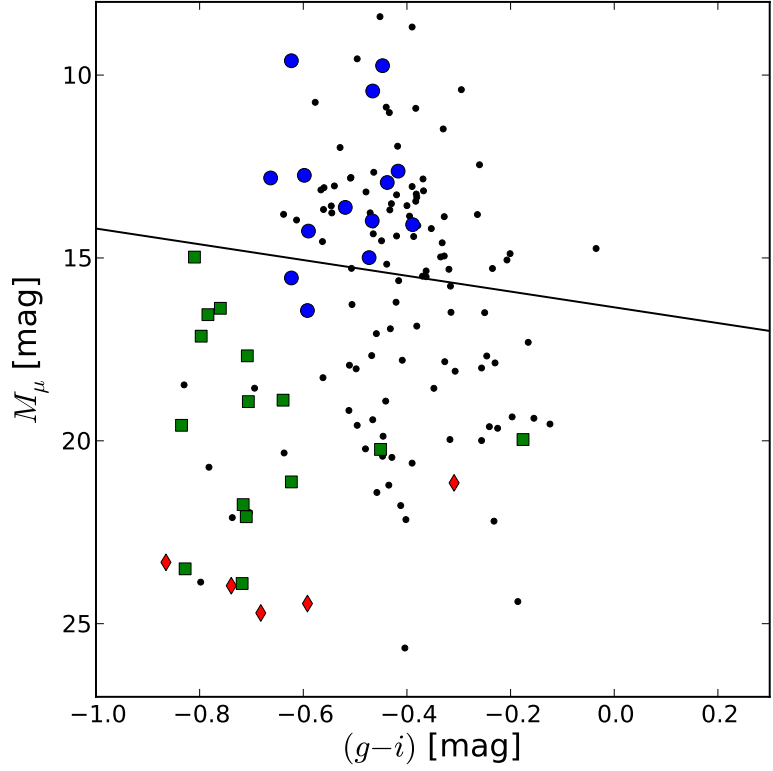}
  \caption{A pseudo HR diagram of the program stars listed in Table~\ref{bstar} using the proper-motion measure $M_\mu$ and $g-i$ color. Spectroscopically classified hot subdwarfs are denoted by squares and white dwarfs by diamonds. Program stars that have not been spectroscopically classified are denoted by small black circles. The solid line separates main-sequence B stars from hot subdwarfs and is taken from \citet{balona2010}.  This figure  provides an excellent tool for distinguishing main-sequence B stars from hot subdwarfs and white dwarfs. }
  \label{fig:hrd}
\end{figure}

The above procedure was further refined by examining the position of a star in a pseudo HR diagram. The construction of this diagram is described in the \citet{balona2010} paper so we only provide an overview of it here.  Using a star's proper motion as a proxy for its distance, a pseudo absolute magnitude is computed from the relation: $M_\mu = g + 5\log(\overline{\mu})$,  where $g$ is the Sloan magnitude listed in the KIC,  $\overline{\mu} =  \Delta\mu/(1~{\rm mas\,yr^{-1}})$, and $\Delta\mu$ is the difference between the stars's proper motion and the center of the B-star concentration.   This absolute magnitude is then plotted against the star's $g-i$ color. Using stars of known MK spectral type, \citet{balona2010} were able to identify a boundary between main-sequence B stars and other stellar types in this pseudo HR diagram.  Since our sample is an extension of the one used in their study, we adopt this same boundary in our study. Figure~\ref{fig:hrd} shows the pseudo HR diagram of our stars. Stars with known spectral types are given special symbols that are discussed in the figure caption. A major difference between our figure and its counterpart in \citet{balona2010} is the number of subdwarfs redder than $g-i = -0.4$. This difference arises from the presence of a sizable number of A stars that were included in our study.  As discussed earlier, the KIC hot star effective temperatures are not very precise. Therefore, to capture as many B stars as possible, our sample extended to stars that have KIC colors appropriate to the early A stars.   Figure \ref{fig:hrd} provides a natural way of separating main-sequence B stars from hot subdwarfs and white dwarfs.

To make our results compatible to those obtained by \citet{balona2010}, we classified stars using the same scheme suggested by those investigators. $\beta$~Cep stars are defined as stars whose pulsation frequencies extend from about 3.5 to 20 d$^{-1}$.  SPB stars are defined as B stars that pulsate with frequencies between 0.5 and 3.5 d$^{-1}$. To be classified as a $\beta$~Cep/SPB hybrid, a B star must possess modes of both of these classes, i.e. between 0.5 to 20 d$^{-1}$. The power spectra of stars with frequency groupings show distinct groupings of many modes around specific frequencies in their power spectra.  Stars with variations due to binarity or rotation show relatively smooth light curves where the variability arises from rotation or proximity effects in a binary system such as light reflection, deformation of the components, or  Doppler beaming. For completeness hot subdwarf and white dwarf candidates are also included in Table~\ref{varclass}.

Table~\ref{varclass} shows the classification assigned to each of our program stars. We assign the columns to KIC  number, quarters observed, our variable classification, the frequency of the waveform with the largest amplitude ($\nu_{\rm max}$), the amplitude of this frequency ($A_{\rm max}$), number of waveforms detected with a amplitude greater the 20 percent of the largest amplitude ($N_{\rm freq}$), and associated notes. In cases where it was not
possible to distinguish a real variation from the low frequency noise, the frequency information is left blank.
When the value for $N_{\rm freq}$ has a colon attached to it, this uncertainty is associated with this noise. The presence of a frequency grouping is denoted in the notes column by the  ``Fg''. When a star's proper motion is large, this is also noted since it is used to differentiate variable classes. Representative members of the each stellar group are discussed below. 

\subsection{$\beta$ Cephei stars}

\begin{figure}
  \centering
  \includegraphics[width=\textwidth]{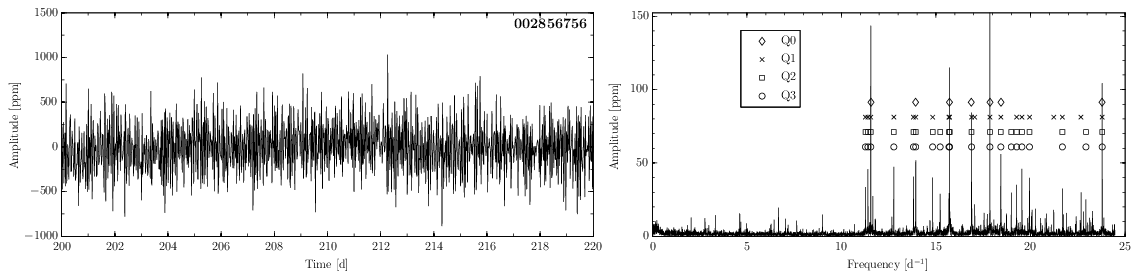}
  \includegraphics[width=\textwidth]{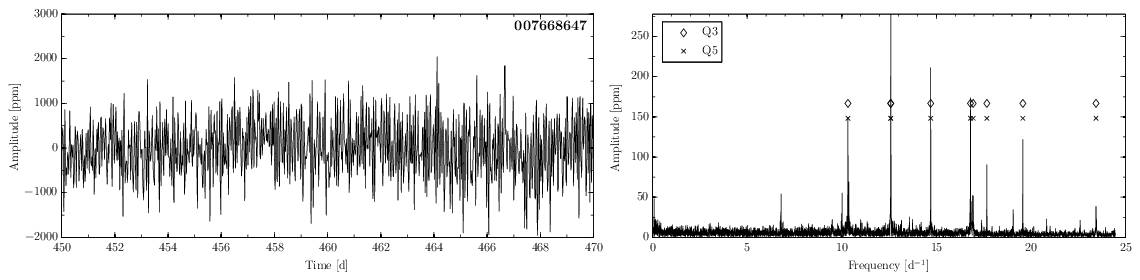}
  \caption{$\beta$~Cep candidate stars. A portion of the Kepler light curves for KIC 002856756 (top) and KIC 007668647 (bottom) and their associated amplitude spectra are shown. The quarters for which data were available and analyzed for each star is labeled at the top of the light curves. Symbols attached to the major peaks in the power spectra show the quarters in which the frequency was detected. To avoid confusion, these symbols are only shown for the frequencies that have an amplitude within 20 percent of the maximum amplitude.}
  \label{fig:bceph}
\end{figure}

The Kepler power spectra of these stars generally possess dozens of frequencies above about 3.5~d$^{-1}$. Few, if any, of these pulsations would have been detected in ground-based studies because of their small amplitudes.  Their spectra often resemble those of hot sub-dwarfs. Therefore, it is possible that some hot subdwarfs have been incorrectly classified as $\beta$~Cep stars.  To minimize this possibility, a candidate star's position in Figures~\ref{fig:vpd} and \ref{fig:hrd}   were used to distinguish hot sub-dwarfs/white dwarfs from B stars. The power spectra of $\beta$~Cep stars can also resemble that of  a $\delta$~Scuti star. In addition, as shown by Figure~\ref{fig:temp}, the Kepler KIC effective temperatures for hot stars are generally much lower than a star's spectroscopically derived temperature. Therefore, stars whose Kepler temperatures would normally be too low to be considered a B star were included in our study. Both of these features make it likely that some of our B-star candidates are actually early A-type stars. The number of $\beta$~Cep stars known prior to 2004 is about 93 \citep{stankov2005}.  We were able to identify 10 additional $\beta$~Cep candidates. In Figure~\ref{fig:bceph}, we provide the amplitude spectra, light curves and stability maps of two of these stars. The stability of the detected oscillations from quarter to quarter is useful in identifying possible transient signals. Additional information about these stars is provided below.

KIC 002856756: In the KIC this star has a Kepler magnitude of 10.2 and has an effective temperature of 10333~K. Its amplitude spectrum, based on the data collected during quarters 0-3 is shown seen  Figure~\ref{fig:bceph}. Its strongest pulsation mode has a SNR of 69 and is located at 17.85 d$^{-1}$ (or a period of 1.34 hours). Although a prominent feature in the amplitude spectrum, the amplitude of this mode is only 0.15 mmag. An examination of its frequency spectrum shows that 19 frequencies have an amplitude of 20 percent of the maximum value. The stronger modes appear to be stable over the time period covered by the Kepler measurements, but this may not be the case for some of the weaker modes. Two frequencies located near 22~d$^{-1}$ were only detected in the Q1 data. Additional measurements are needed to investigate whether some weaker modes are actually transient or are the result of instrumental effects. Although we are using this star are an example of a $\beta$~Cep star, their is some evidence for several small amplitude (less than 20~ppm) modes with frequencies of less than 3.5~d$^{-1}$. If confirmed, this star would then be a hybrid. One wonders if classically defined $\beta$~Cep stars were re-examined using high S/N data sets like those available from Kepler, whether many of them would be re-classified as hybrids.
 
KIC 007668647: In the KIC this star has a Kepler magnitude of 15.4 and an effective temperature of 10668 K. Its strongest mode has a frequency of 12.59~d$^{-1}$ and an amplitude of 0.29~mmag. Twenty-four frequencies were detected within 20 percent of this maximum amplitude. Many of the detected modes are closely spaced in frequency and are not obvious in Figure~\ref{fig:bceph}. Some of these may be artifacts of this star's relatively complex window function. All of the stronger frequencies found in this star appear to be stable over the time  period covered by Kepler Q3 and Q5.

\subsection{SPB Stars }

\begin{figure}
  \centering
  \includegraphics[width=\textwidth]{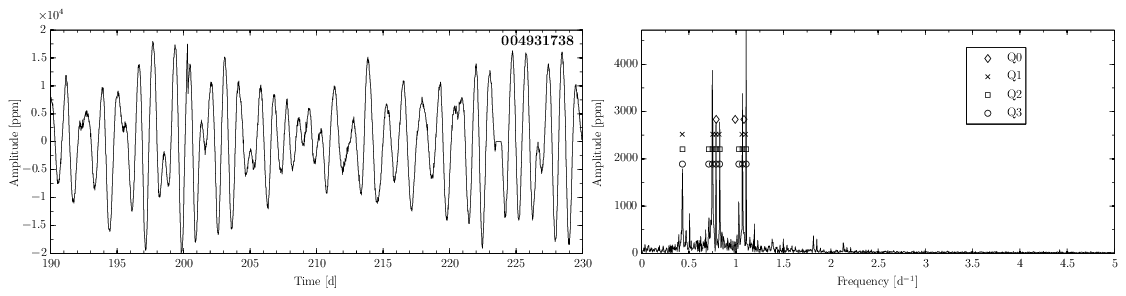}
  \includegraphics[width=\textwidth]{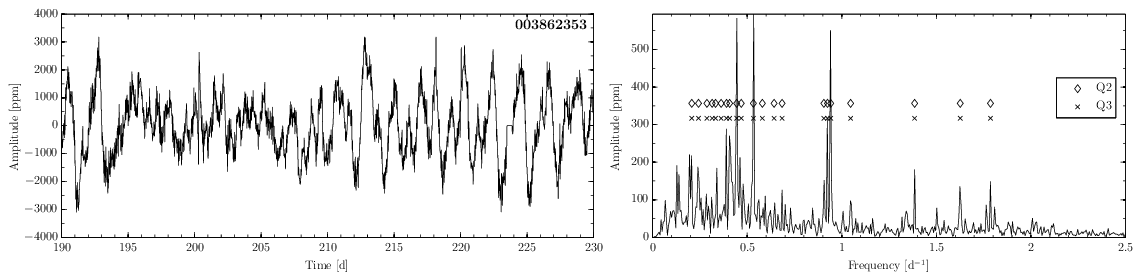}
  \caption{SPB candidate stars.  A portion of the Kepler light curves for KIC 004931738 (top) and KIC 003862353 (bottom) and their associated amplitude spectra are shown. The quarters for which data were available and analyzed for each star is labeled at the top of the light curves. Symbols attached to the major peaks in the power spectra show the quarters in which the frequency was detected. To avoid confusion, these symbols are only shown for the frequencies that have an amplitude within 20 percent of the maximum amplitude.}
  \label{fig:spb}
\end{figure}

To be classified as an SPB a star had to meet three criteria: (1) all of the frequencies in its power spectrum had to be $\le 3.5\,{\rm d^{-1}}$;  (2) its positions in Figures~\ref{fig:vpd} and \ref{fig:hrd} had to be consistent with that of a B star; and (3) its power spectrum had to contain more than two frequencies.  The last criterion was imposed to distinguish SPB stars from stars classified as binary/rotation (Bin/Rot). It was assumed that frequencies arising from rotation or binary activity will be few in number. Although special cases may arise where this assumption is not valid, it conforms to the scheme used to classify stars in the \citet{balona2010} study. Approximately 51 SPBs were known to exist prior to about 2007 \citep{decat2007}. We identified an additional 46 potential members of this class. Example amplitude spectra, light curves, and stability maps of two members of this group are shown in Figure~\ref{fig:spb} and are discussed below. 

KIC 004931738: This star has a Kepler magnitude of 11.6 and KIC effective temperature of 10136 K. Its light curve extends for approximately 225 days and includes measurements made during quarters 0-3.  Its largest amplitude signal (5.3 mmag) is at a frequency of 0.75~d$^{-1}$. Seven other frequencies have amplitudes that are within 20 percent of this peak amplitude.  Frequencies are present out to at least 1.8~d$^{-1}$. The  spectrum consists of three frequency groupings, separated by about 0.31 d$^{-1}$, centered at 0.43 d$^{-1}$, 0.75 d$^{-1}$, and 1.06  d$^{-1}$. The strongest frequencies within these groups are separated by about 0.038~d$^{-1}$. Figure~\ref{fig:spb} shows that the stronger amplitude frequencies are present in each of the quarters for which we have Kepler data, with the exception of quarter 0 which was only 10 days long, and thus appear to be quite stable.

KIC 003862353: This star has a Kepler magnitude of 14.4 and a KIC listed effective temperature of 10738 K. Measurements were obtained during quarters 2 and 3 and cover about 180 days. The strongest mode has a frequency of 0.44~d$^{-1}$ and an amplitude of 0.65~mmag. Smaller amplitude modes are present out to a frequency of at least 2.8~d$^{-1}$.  Its spectrum  in Figure~\ref{fig:spb} is more complex than for KIC 4931738, but no regular frequency spacings are evident. Eighteen frequencies were found that have an amplitude at least 20 percent as strong as that of the primary mode.

\subsection{SPB/$\beta$~Cep Hybrid Stars}
\begin{figure}
  \centering
  \includegraphics[width=\textwidth]{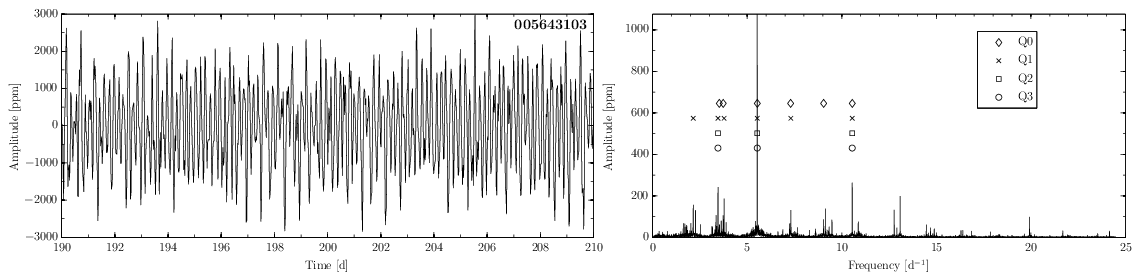}
  \includegraphics[width=\textwidth]{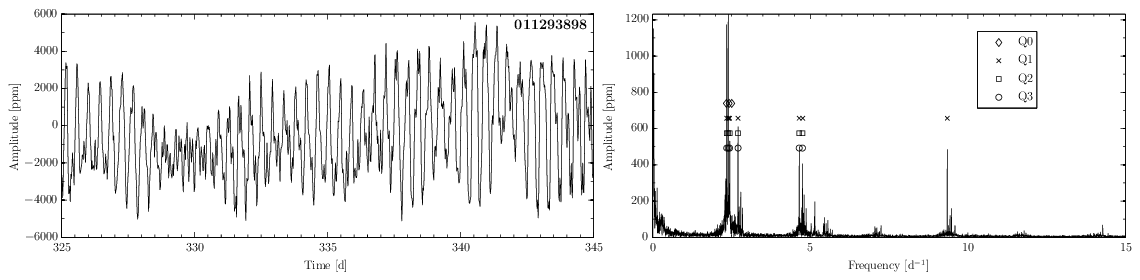}
  \caption{$\beta$~Cep/SPB hybrid candidate stars. A portion of the Kepler light curves for KIC 005643103 (top) and KIC 011293898 (bottom) and their associated amplitude spectra are shown. The quarters for which data were available and analyzed for each star is labeled at the top of the light curves. Symbols attached to the major peaks in the power spectra show the quarters in which the frequency was detected. To avoid confusion, these symbols are only shown for the frequencies that have an amplitude within 20 percent of the maximum amplitude.}
  \label{fig:hyb}
\end{figure}

Hybrid stars pulsate with frequencies typical of both the SPB stars and $\beta$~Cep stars, i.e., within the frequency range 0.5 to 20~d$^{-1}$. About a third of the stars we identified in this group possess frequency groupings. Prior to about 2007 only about 6 hybrids were known \citep{decat2007}. Fourteen stars in our sample are classified as hybrid pulsators. The light curves, amplitude spectra, and stability maps of two of them are shown in Figure~\ref{fig:hyb}. Some additional discussion of these stars is provided below.

KIC 005643103 ($\beta$~Cep/SPB): The star was observed by Kepler for quarters 0-3, providing nearly 230 days of continuous measurements. Its Kepler magnitude and effective temperature are 9.74 and 10225~K, respectively. Its primary frequency at 5.53~d$^{-1}$ has an amplitude of 1.1~mmag.  The star has power at frequencies extending from  1.6~d$^{-1}$ to about 22~d$^{-1}$ although only 3 other modes have amplitudes greater than 20\% that of its primary mode. Its power spectrum possesses several frequency groupings, but their spacing is not even.   Two groupings near 7~d$^{-1}$ were not detected in quarters 2 and 3.

KIC 011293898 (SPB/$\beta$~Cep). This star has a Kepler magnitude of 13.4 and a KIC effective temperature of 15072~K. Measurements were obtained in quarters 0-3 and cover about 230 days. The highest peak in the amplitude spectrum has a frequency of 2.35~d$^{-1}$ and an amplitude of 1.49~mmag. Approximately 17 other frequencies have an amplitude that is within 20 percent of this value. Frequency groupings starting at 2.35~d$^{-1}$ and spaced at integer multiples of this frequency are present. The pattern of the power within these frequency groupings, however, differs. The higher amplitude modes are stable over our observing period for this star, but at least one low amplitude mode near 5~d$^{-1}$ was  detected only in quarter 0. The frequency grouping near 10~d$^{-1}$ was only detected in quarter 1.

\subsection{Binary/Rotation modulated stars}
\begin{figure}
  \centering
  \includegraphics[width=\textwidth]{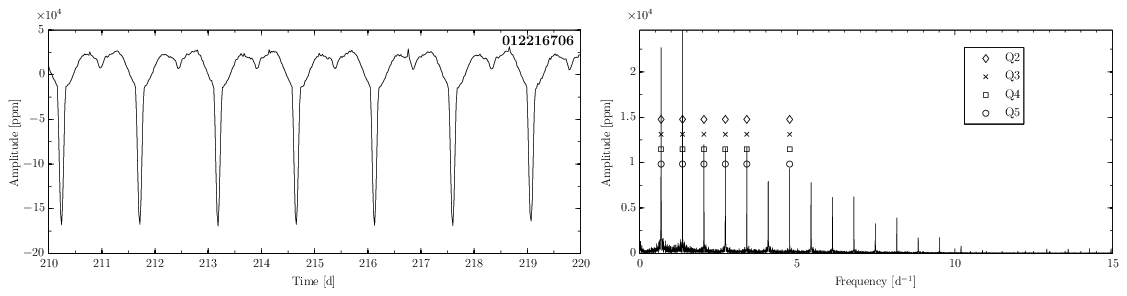}
  \includegraphics[width=\textwidth]{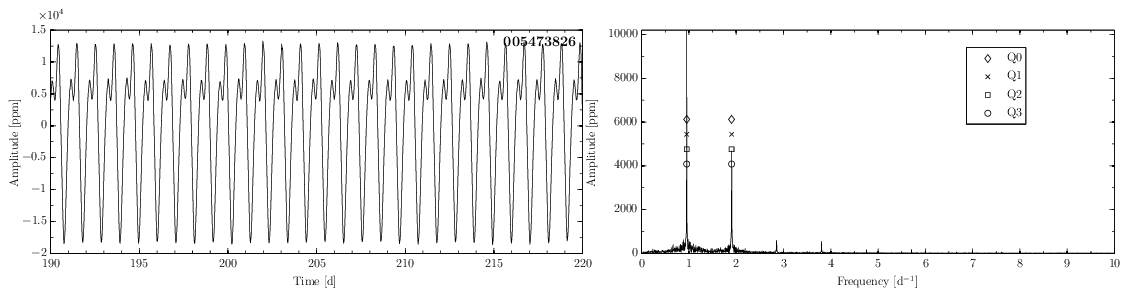}
  \caption{Bin/Rot candidate stars. A portion of the Kepler light curves for KIC 012216706 (top) and KIC 005473826 (bottom) and their associated amplitude spectra are shown. The quarters for which data were available and analyzed for each star is labeled at the top of the light curves. Symbols attached to the major peaks in the power spectra show the quarters in which the frequency was detected. To avoid confusion, these symbols are only shown for the frequencies that have an amplitude within 20 percent of the maximum amplitude.}
  \label{fig:bin}
\end{figure}

The light curves of stars within this group vary smoothly. Their power spectra show few peaks and these are thought to be connected to the star's rotation or binarity rather than pulsation.  We identified 67 members of this group. Eight of these stars have high surface gravities and two have A-star effective temperatures in the KIC.  The light curves, amplitude spectra, and stability maps of two of these stars are shown in Figure~\ref{fig:bin}. Each is briefly described below.

KIC 012216706: The Kepler magnitude and effective temperature of this star are 10.7 and 11061~K. Its light curve is typical of a binary system of one large star and one small companion. The strongest mode present in the power spectrum has a frequency and amplitude of  0.53~d$^{-1}$ and 0.11 magnitudes. Other frequencies in the power spectrum are integer multiples of this primary frequency. This frequency is stable over the time period of our data set which consists of quarters 0-2.

KIC 005473826: This star has a Kepler magnitude of 10.9 and a KIC effective temperature of 10927~K.  Its light curve is smoothly varying and its power spectrum shows only two dominant modes at 0.95~d$^{-1}$ and 1.90~d$^{-1}$. The latter is simply a harmonic of the first frequency, indicating that this waveform has a non-sinusoidal shape. Both of these frequencies, which have amplitudes of 10.80~mmag and 6.03~mmag respectively, are stable over the time period (quarters 0-3) covered by our data. The light curve is suggestive of magnetic spot modulation.

\subsection{Pulsating Hot Subdwarfs and White Dwarfs}
Candidates for these two classes of stars are presented because of their astrophysical importance. The power spectra of hot horizontal branch stars provide information about the structure of evolved stars, while white dwarfs are a major end product of stellar evolution. Stars are classified as 
pulsating hot subdwarfs or pulsating white dwarfs based on either their location in Figure~\ref{fig:hrd} or their spectroscopic classification \citep{ostensen2010}. Membership in either of these classes is highly uncertain. Many of these stars may be misclassified $\beta$~Cep or SPB stars.  Generally they possess a large proper motion and nearly all of the pulsating white dwarf candidates have a KIC log(g) $\>$ 5. In Table~\ref{varclass} we classify 12 stars as pulsating sdB stars and 11 stars as pulsating white dwarfs.  The discovery of pulsating white dwafts in the Kepler field is interesting since only a few (one?) of these stars are currently known.  Many of the white dwarf candidates possess frequency groupings.  For two sdB stars, KIC 6848529 and 9543660, we do not provide frequencies. In both cases the light curves are affected by unresolved calibration issues that cause a rapid rise in power at low frequencies. It is likely that the star KIC 006848529 has a stable mode at 0.92~d$^{-1}$ with an amplitude of about 0.13~mmag,  but a cleaner data set is needed to confirm this value. The amplitude spectrum of KIC 009543660 has several peaks below 1~d$^{-1}$, but their amplitudes are very small ($\le 0.03$~mmag) and a confirming data set is required.

\section{Summary}
\label{summ}

In this study we have examined the light curves of  252 B-star candidates in the Kepler catalogue. Variable stars in this sample have been classified using the criteria adopted by \citet{balona2010}.  These groupings include:  the $\beta$~Cep stars, the SPB stars, hybrid pulsators whose power spectra contain frequencies present in both of these pulsators, and B stars whose variability is likely related to their rotation or presence in a binary. Proper motions obtained from the UCAC3 catalogue \citep{zacharias2009} and the position of these stars in a pseudo HR diagram were used to distinguish between likely main-sequence B stars, hot subdwarfs, and A stars. Because the Kepler temperatures of hot stars are  poorly known, the stars classified in this study should only be considered to be candidates for each of their variable classes. The primary result of this study has been to increase the number of B star variables identified using Kepler light curves. Using \citet{balona2010} as a reference and the variable  candidates studied here for inclusion within their respective classes, the number of $\beta$~Cep candidates has been increased from 0 to 10, the number of SPB stars from 8 to 54, the number of hybrids from 7 to 21, the number of Bin/Rot B stars from 23 to 82. For comparison purposes the number of confirmed SPBs and hybrids known prior to 2007 were about 51 and 6 respectively \citep{decat2007}. The number of confirmed $\beta$~Cep stars know prior to 2004 was 93 \citep{stankov2005}.  A secondary result of this study is the identification of an additional 11 white dwarf candidates, four of which possess frequency groupings.

\acknowledgments

We acknowledge support from the \textit{Kepler Guest Observer Program}  and Cycle 1 project GO100003. Part of this work was also funded by the NASA EPSCoR award to NMSU under contract \#~NNX09AP76A and NSF PAARE award AST-0849986. Fruitful discussions about this work with Luis Balona, Andrzej Pigulski, Thomas Harrison, Patrick Gaulme, and Catherine Lovekin are acknowledged.

\bibliographystyle{abbrvnat}
\bibliography{myrefs}

\begin{deluxetable}{lllll}
\tablecaption{KIC numbers of non--variable stars in sample.   \label{nonvar} }
\tablewidth{0pt}
\tabletypesize{\small}
\startdata
1028085	&	4678193	&	6761643	&	8758259	&	10685044*	\\
1721361	&	4829241	&	6853881	&	9005047*&	10789011	\\
2020175	&	5109770	&	6860325*&	9032986*&	10816401	\\
2308039	&	5288186	&	6946179*&	9141584*&	11075420*	\\
2581243	&	5372254	&	6950240	&	9151452	&	11351218	\\
2830941	&	5544772*&	7104168	&	9718641	&	11395846	\\
2987372	&	5557097	&	7188447	&	9718683	&	11403244	\\
3112242	&	5612463*&	7353409	&	9725496	&	11456087	\\
3343613	&	5951261*&	7358074*&	9826266	&	11565279	\\
3345780	&	5960374	&	7549771	&	9948188	&	11652092*	\\
3353239	&	6204719	&	7918401*&	9957741	&	11713245*	\\
3442422	&	6205507	&	7966886	&	10272154&	11719437	\\
3459901	&	6214434	&	8037519	&	10292000&	11719549	\\
3544662	&	6267975	&	8077281	&	10340843&	11817929$^\dagger$	\\
3545048	&	6293167	&	8167479	&	10389280*&	11822535	\\
3858566*&	6371916	&	8442729*&	10415184&	12060050	\\
4166305	&	6391844*&	8523491	&	10449976&	12066844	\\
4366757*&	6460385*&	8559392	&	10483888&	12366417	\\
4457244	&	6688779	&	8609006	&	10535643*&	12453485*	\\
4660027	&	6752816	&	8682822	&	10671396&	12885346	
\enddata
\tablenotetext{*}{Denotes that pulsations might be present but are below the 20~ppm level.}
\tablenotetext{\dagger}{Denotes stars in common with \citet{balona2010}.}
\end{deluxetable}

\begin{deluxetable}{l l l l l l l}
\tablecaption{B-type star target list with general parameters. \label{bstar}}
\tablewidth{0pt}
\tabletypesize{\small}
\tablehead{
\colhead{KIC \#} & \colhead{$K_p$} & \colhead{$g$} & \colhead{$g-i$} & \colhead{$\mu_{\alpha}\cos\delta$} & \colhead{$\mu_{\delta}$} & \colhead{$M_{\mu}$}\\
\colhead{} & \colhead{} & \colhead{} & \colhead{} & \colhead{[mas/yr]} & \colhead{[mas/yr]} & \colhead{}
}

\startdata
1430353	&	12.39	&	12.320	&	-0.26	&	-3.5	&	-4.7	&	12.45	\\
1718290	&	15.49	&	15.323	&	-0.43	&	-6.2	&	-6.3	&	16.94	\\
1868650	&	13.45	&	13.453	&	-0.71	&	4.7	&	-14	&	18.93	\\
2697388	&	15.39	&	15.130	&	-0.64	&	0.9	&	-7.6	&	18.89	\\
2708156	&	10.67	&	10.672	&	-0.04	&	1.9	&	-7.4	&	14.74	\\
2853320	&	13.71	&	13.581	&	-0.37	&	-4	&	-3	&	15.5		\\
2856756	&	10.25	&	10.163	&	-0.26	&	-1.7	&	-0.7	&	13.81	\\
2860851	&	10.11	&	9.996	&	-0.3	         &	-4.2	&	-4.2	&	10.4		\\
2987640	&	12.64  	&	12.500	&	-0.44	&	-1.6	&	-3.3	&	15.17	\\
3216449	&	11.82	&	11.670	&	-0.43	&	-6.5	&	-6.2	&	13.52	\\
3443222	&	14.33	&	14.291	&	-0.17	&	-4.6	&	-9.4	&	17.31	\\
3443582	&	15.8	         &	15.583	&	-0.51	&	-9.4	&	-4.3	&	19.17	\\
3459297	&	12.55	&	12.424	&	-0.35	&	-5.9	&	-3.8	&	14.2		\\
3527751	&	14.86	&	14.567	&	-0.71	&	-1.1	&	-8.1	&	17.68	\\
3544662	&	15.03	&	14.878	&	-0.41	&	-4.8	&	18.5	&	21.77	\\
3561700	&	14.91	&	14.864	&	-0.19	&	37.5	&	-74.3&	24.4		\\
3756031$^\dagger$	&	10.04	&	9.860	&	-0.47	&	-5.6	&	-5.5	&	10.44	\\
3763002	&	13.69	&	13.607	&	-0.25	&	-4.9	&	-11.9   &	17.68	\\
3839930$^\dagger$	&	10.86	&	10.675	&	-0.51	&	-6.7	&	-6.6	&	12.82	\\
3862353 &       14.43   &       14.440  &       -0.13   &               &               &               \\
3865742$^\dagger$	&	11.13	&	10.963	&	-0.43	&	-3.8	&	-4.5	&	11.03	\\
4064365	&	12.3	         &	12.167	&	-0.33	&	-3.1	&	-2.6	&	14.59	\\
4077252	&	12.31	&	12.251	&	-0.21	&	-0.9	&	-4.1	&	15.06	\\
4373195 &       9.12    &       8.940   &       -0.45   &       -4.8    &      -4.8     &      8.40      \\
4373805	&	13.54	&	13.433	&	-0.32	&	-8.2	&	-4.2	&	16.49	\\
4476114	&	12.92	&	12.758	&	-0.42	&	-7.4	&	-1.6	&	16.21	\\
4547333	&	16.32	&	16.167	&	-0.40	         &	7.4	&	73.1	&	25.67	\\
4663658 &       11.32   &       11.199  &       -0.32   &       0.3 & 1.4 & 15.77      \\
4828345	&	12.78	&	12.597	&	-0.47	&	-1	&	4.4	&	17.67	\\
4831616 & 14.77 & 14.820 & -0.21 & & & \\
4931738	&	11.65	&	11.501	&	-0.38	&	-4	&	-4.7	&	10.91	\\
4936089	&	11.93	&	11.760	&	-0.37	&	-5.6	&	-6.8	&	13.17	\\
4939281	&	12.08	&	11.939	&	-0.39	&	-4.2	&	-5.6	&	8.69		\\
5084439	&	14.36	&	14.242	&	-0.31	&	-4.1	&	0.5	&	18.1		\\
5172153	&	14.62	&	14.552	&	-0.24	&	-4.2	&	-6.8	&	15.29	\\
5352328 & 15.01  &  14.853  &  -0.39  &  -12.3  &  6.3  &  20.61 \\
5450881	&	12.47	&	12.329	&	-0.38	&	-5.8	&	-5.1	&	13.25	\\
5473826	&	10.89	&	10.753	&	-0.38	&	-3.2	&	-8.5	&	13.34	\\
5477601	&	12.79	&	12.732	&	-0.2	         &	-2.9	&	-3.1	&	14.88	\\
5479821$^\dagger$	&	9.89	         &	9.743	&	-0.42	&	-2.3	&	-8.6	&	12.63	\\
5530935	&	8.04	         &	7.833	&	-0.55	&	7.3	&	2.6	&	13.58	\\
5643103	&	9.74	         &	9.565	&	-0.33	&	-2.7	&	-7.2	&	11.47	\\
5687841 &  10.39 &  10.213 & -0.42 & -2.5 &  -6.7 & 11.95 \\
5706079	&	11.58	&	11.447	&	-0.37	&	-4.3	&	-3.5	&	12.84	\\
5775128	&	13.54	&	13.319	&	-0.55	&	-7.8	&	5.7	&	18.65	\\
5807616	&	15.02	&	14.731	&	-0.72	&	-2.4	&	-30.6&	21.74	\\
5941844	&	9.28	         &	9.050	&	-0.56	&	2.2	&	5.4	&	14.55	\\
5942605	&	14.08	&	13.868	&	-0.51	&	-14.7&	2.5	&	19.45	\\
6188286	&	14.21	&	13.855	&	-0.81	&	-5.9	&	-5.9	&	14.98	\\
6278403	&	8.76	         &	8.541	&	-0.55	&	4.4	&	1.5	&	13.77	\\
6363494	&	14.42	&	14.341	&	-0.26	&	-4.6	&	8.1	&	19.99	\\
6614501 &  16.09 & 16.19 & -0.26 & & & \\
6777127	&	14.23	&	14.145	&	-0.24	&	-15.7&	-0.5	&	19.61	\\
6780397	&	10.05	&	9.821	&	-0.54	&	-8.3	&	-7.2	&	13.03	\\
6848529	&	10.73	&	10.407	&	-0.78	&	4.1	&	-20.1&	16.55	\\
6864569	&	10.05	&	9.871	&	-0.48	&	-0.7	&	-2.5	&	13.2		\\
6866662	&	13.6	         &	13.425	&	-0.47	&	-5.4	&	-5	&	13.77	\\
6878288 &    16.67 & 16.75 & -0.29  & & & \\ 
6950556	&	12.75	&	12.629	&	-0.33	&	-6	&	-4.9	&	13.87	\\
6953047 &  14.31 & 14.34  & -0.15 & & & \\
6954726$^\dagger$	&	11.93	&	11.715	&	-0.52	&	-4.2	&	-3	&	13.62	\\
7022658	&	14.53	&	14.451	&	-0.26	&	-6.8	&	-9.9	&	18.01	\\
7098125	&	14.68	&	14.505	&	-0.46	&	-23.2&	9.5	&	21.41	\\
7108883	&	13.36	&	13.241	&	-0.33	&	-4.4	&	-13.7&	17.84	\\
7204794	&	14.16	&	14.092	&	-0.25	&	-7.3	&	-5.8	&	16.5		\\
7282265	&	14.29	&	14.179	&	-0.32	&	-10.4&	7.6	&	19.96	\\
7335517	&	15.75	&	15.410	&	-0.8		&	-53.1&	-0.3	&	23.86	\\
7353409	&	14.68	&	14.317	&	-0.84	&	-9.3	&	-15.5&	19.58	\\
7434250	&	15.47	&	15.278	&	-0.45	&	1.9	&	-13	&	20.24	\\
7540755	&	14.23	&	14.083	&	-0.35	&	-7	&	2	&	18.56	\\
7599132$^\dagger$	&	9.39	         &	9.220	&	-0.44	&	3.9	&	7	&	15.08	\\
7630417	&	13.79	&	13.716	&	-0.23	&	-11	&	-6.4	&	17.87	\\
7664467 &  16.45  &  16.53 & -0.31 & & & \\
7668647	&	15.4	         &	15.058	&	-0.84	&	-10.2&	0.1	&	19.59	\\
7749504	&	12.72	&	12.551	&	-0.47	&	-4.7	&	-3.4	&	14.1		\\
7755741	&	13.75	&	13.402	&	-0.8		&	-1.8	&	-10.4&	17.14	\\
7778838	&	11.87	&	11.758	&	-0.33	&	-3.7	&	-9.7	&	14.95	\\
8054179	&	14.43	&	14.127	&	-0.7		&	1.7	&	31	&	21.96	\\
8087269$^\dagger$	&	12.13	&	11.943	&	-0.47	&	-7.9	&	-7.3	&	14.99	\\
8129619	&	10.99	&	10.833	&	-0.43	&	-2.4	&	-8.6	&	13.69	\\
8161798$^\dagger$	&	10.47	&	10.329	&	-0.44	&	-3.4	&	-8.6	&	12.94	\\
8167938	&	10.85	&	10.668	&	-0.46	&	-1.9	&	-4.7	&	12.66	\\
8183197	&	11.22	&	11.093	&	-0.36	&	-0.8	&	0.8	&	15.36	\\
8233804	&	15.28	&	15.109	&	-0.44	&	-7.5	&	-0.6	&	18.91	\\
8264293	&	11.31	&	11.166	&	-0.38	&	-3.1	&	-8	&	13.45	\\
8302197	&	16.43	&	16.165	&	-0.64	&	-10.6&	-2.8	&	20.33	\\
8362546	&	15.69	&	15.520	&	-0.44	&	9.1	&	-8.5	&	21.21	\\
8488717$^\dagger$	&	11.76	&	11.616	&	-0.39	&	-6.3	&	-3	&	14.09	\\
8619436	&	11.81	&	11.666	&	-0.38	&	-3	&	-2.6	&	14.11	\\
8619526	&	15.84	&	15.493	&	-0.74	&	-34.1&	-44.8&	23.96	\\
8754603	&	14.16	&	13.991	&	-0.45	&	-8.8	&	13.4	&	20.42	\\
8759258	&	12.42	&	12.275	&	-0.38	&	-10.1&	-11.3&	16.86	\\
8766405$^\dagger$	&	8.88	         &	8.706	&	-0.45	&	-4.1	&	-7	&	9.74		\\
8871494	&	13.72	&	13.520	&	-0.51	&	-8.3	&	1.1	&	17.93	\\
9020774	&	15.09	&	14.910	&	-0.48	&	-12.1&	3.1	&	20.22	\\
9075895	&	14.63	&	14.449	&	-0.47	&	-6.2	&	4.3	&	19.42	\\
9202990	&	15.04	&	14.904	&	-0.31	&	-20.1&	2.7	&	21.15	\\
9211123	&	16.1	         &	15.895	&	-0.4	         &	11.4	&	3.1	&	22.15	\\
9227988	&	12.85	&	12.731	&	-0.36	&	-4.4	&	-1.8	&	15.51	\\
9278405	&	10.24	&	10.024	&	-0.5	         &	-3.6	&	-5.8	&	9.56		\\
9468611	&	13.14	&	12.990	&	-0.42	&	-4	&	-4.3	&	13.27	\\
9472174	&	12.26	&	11.964	&	-0.76	&	2.8	&	-2.6	&	16.38	\\
9543660	&	13.77	&	13.475	&	-0.72	&	-63.4&	101	&	23.9		\\
9582169	&	12.05	&	11.873	&	-0.44	&	-4.5	&	-6	&	10.88	\\
9663239	&	14.2	         &	14.138	&	-0.23	&	-3.7	&	35.5	&	22.2		\\
9715163	&	8.28	         &	8.063	&	-0.56	&	4.5	&	-0.5	&	13.08	\\
9715425	&	13.07	&	12.911	&	-0.42	&	-5.6	&	-3.9	&	14.4		\\
9884329	&	12.14	&	11.991	&	-0.39	&	-1.7	&	-3.8	&	14.41	\\
9910544	&	10.84	&	10.695	&	-0.39	&	-5.5	&	-8.1	&	13.05	\\
9958053	&	11.73	&	11.603	&	-0.34	&	-0.5	&	-2.6	&	14.97	\\
10090722	&	13	         &	12.867	&	-0.32	&	-6.1	&	-2.9	&	15.31	\\
10118750	&	13.9	         &	13.698	&	-0.5	         &	-8.1	&	0.9	&	18.03	\\
10130954$^\dagger$	&	11.13	&	10.850	&	-0.66	&	-3	&	-7.5	&	12.81	\\
10139564	&	16.13	&	15.759	&	-0.83	&	10.9	&	26.5	&	23.5		\\
10207025	&	15.04	&	14.694	&	-0.78	&	11.3	&	-9.2	&	20.72	\\
10220209	&	14.12	&	13.882	&	-0.59	&	1.5	&	12.5	&	20.25	\\
10281890	&	10.41	&	10.268	&	-0.4	         &	0.7	&	-6.9	&	13.86	\\
10285114$^\dagger$	&	11.23	&	10.991	&	-0.59	&	-6.2	&	-1.3	&	14.27	\\
10526294	&	13.03	&	12.834	&	-0.51	&	-3.4	&	-5.8	&	12.8		\\
10536147$^\dagger$	&	11.59	&	11.333	&	-0.64	&	-4.7	&	-2.3	&	13.81	\\
10551350	&	14.52	&	14.502	&	-0.12	&	-10.6&	-13.4&	19.54	\\
10553698	&	15.13	&	14.837	&	-0.69	&	-1.5	&	-10.2&	18.56	\\
10657664	&	13.23	&	13.082	&	-0.42	&	-2.9	&	-2.5	&	15.62	\\
10658302$^\dagger$	&	13.13	&	12.861	&	-0.62	&	-4.1	&	-5.5	&	9.61		\\
10670103	&	16.53	&	16.280	&	-0.62	&	-5.7	&	-14.6&	21.12	\\
10790075	&	12.81	&	12.597	&	-0.51	&	-1.3	&	-3.7	&	15.28	\\
10797526$^\dagger$	&	8.34	         &	8.086	&	-0.61	&	2.4	&	8	&	13.96	\\
10799291	&	14.98	&	14.801	&	-0.46	&	-2.1	&	-3.6	&	17.07	\\
10904353	&	14.19	&	14.005	&	-0.5	         &	-15	&	-12.8&	19.58	\\
10937527	&	14	         &	13.856	&	-0.41	&	-10.2&	-7.1	&	17.8		\\
10966623	&	14.87	&	14.643	&	-0.56	&	-9.6	&	-5.9	&	18.27	\\
10982905	&	14.15	&	14.093	&	-0.18	&	6.8	&	4.6	&	19.96	\\
11038518	&	13.55	&	13.379	&	-0.45	&	-4.3	&	-7.1	&	14.53	\\
11045341	&	14.87	&	14.828	&	-0.2	         &	3	&	-8.7	&	19.35	\\
11152422	&	15	         &	14.971	&	-0.16	&	-5.7	&	-12.9&	19.38	\\
11287855	&	14.19	&	14.027	&	-0.45	&	-19	&	-3.9	&	19.87	\\
11293898	&	13.45	&	13.239	&	-0.51	&	-4.9	&	-1.4	&	16.27	\\
11302565	&	15.58	&	15.407	&	-0.43	&	-4.9	&	4.8	&	20.45	\\
11360704$^\dagger$	&	10.74	&	10.506	&	-0.56	&	-4.1	&	-1.1	&	13.68	\\
11454304$^\dagger$	&	12.95	&	12.687	&	-0.62	&	-7.8	&	-4.1	&	15.55	\\
11558725	&	14.95	&	14.586	&	-0.83	&	-10.2&	-6.4	&	18.47	\\
11671923	&	10.68	&	10.525	&	-0.42	&	-4.7	&	-2.8	&	12.63	\\
11912716	&	6.6	         &	6.370	&	-0.57	&	0.4	&	16.7	&	13.14	\\
11953267	&	13.53	&	13.302	&	-0.59	&	-2.9	&	-1.4	&	16.44	\\
11957098	&	12.89	&	12.708	&	-0.47	&	-4	&	-7.5	&	14.34	\\
11971405	&	9.32	         &	9.079	&	-0.58	&	-2.3	&	-6.2	&	10.75	\\
12021724	&	15.59	&	15.284	&	-0.74	&	8.2	&	14	&	22.1		\\
12116239	&	15.02	&	14.957	&	-0.23	&	-13	&	-5.8	&	19.66	\\
12207099$^\dagger$	&	10.28	&	10.107	&	-0.47	&	0.3	&	-9.2	&	13.99	\\
12217324$^\dagger$	&	8.31	         &	8.098	&	-0.53	&	1.4	&	-3.6	&	11.98	\\
12258330$^\dagger$	&	9.52	         &	9.249	&	-0.62	&	-20.2&	7.7	&	15.82	\\
12268319	&	11.63	&	11.475	&	-0.4	         &	-2.3	&	-7.1	&	13.57	\\
\enddata
\tablenotetext{\dagger}{Denotes stars in common with \citet{balona2010}.}
\end{deluxetable}

\begin{deluxetable}{rccllll}
\tablecaption{Variable classifications. \label{varclass}}
\tablewidth{\textwidth}
\tabletypesize{\small}
\tablehead{
\colhead{KIC \#} & \colhead{Quarters} & \colhead{Class} & \colhead{$\nu_{\rm max}$} & \colhead{$A_{\rm max}$} & \colhead{$N_{\rm freq}$} & Notes\\
\colhead{} & \colhead{} & \colhead{} & \colhead{[c/d]} & \colhead{[ppm]} & \colhead{} & \colhead{}
}
\startdata
2856756	 & 0123	&  BCEP   & 17.86 &  151 &  19 &	        \\
6614501  & 35   &  BCEP   & 6.35  &  231 &   6&  \\
7204794	 & 23	&  BCEP   &  3.87 & 3222 &   3 & Possible $\delta$ Scuti \\
7668647	 & 235	&  BCEP   &  12.59 & 286 &   24 &	        \\
8302197	 & 35	&  BCEP   & 24.47 &  432 &  10 &               \\
10118750 & 123	&  BCEP   &  3.63 & 1360 &  20 & Fg	        \\
10553698 & 45	&  BCEP   & 17.45 & 1135 &  11 &               \\
10670103 & 25	&  BCEP   & 11.93 & 1858 &  29 &	        \\
11045341 & 2345	&  BCEP   &  7.42 & 3412 &   6 & Possible $\delta$ Scuti \\
11558725 & 35	&  BCEP:  & 11.33 &  330 &  52 & Possible sdB, large $\mu$ \\
\\
1430353	 & 0123	&SPB	&0.98	& 4188   &   7 &		\\
3443222	 & 23	&SPB	&1.10	&  981   &  12 & Possible A star	\\
3459297	 & 0123	&SPB	&0.99	& 4074   &   8 &		\\
3756031$^\dagger$	 & 0123	&SPB	&0.40	&  574   &  40 &		\\
3839930$^\dagger$	 & 123	&SPB	&0.86	& 8999   &   3 &		\\
3862353  &  23  &SPB   &0.44   & 651    &  19 &    \\
3865742$^\dagger$	 & 0123      &SPB	&0.24	& 7783	 &   4 &		\\
4077252	 & 0123	 &SPB	&0.25	& 705	 &  10 &		\\
4373195  & 0123 &SPB  &0.31   &   18   &  12 & Noisy low freq. spectrum  \\
4663658  & 0123 &SPB  &1.32   &    9   &  11 & Fg\\
4931738	 & 0123	&SPB	&0.75	& 5311   &   8 &		\\
4936089	 & 123	&SPB	&0.87	& 5391   &   5 &		\\
4939281  & 0123	&SPB    & 0.27  & 1126   &   9 &Possible Hybrid	\\
5084439	 & 23	&SPB	&1.71	& 3492   &  16 &Fg		\\
5477601	 & 023	&SPB	&0.19	& 320 	 &  12 &		\\
5941844	 & 0123	&SPB	&1.31	& 12994  &   4 &		\\
6780397	 & 0123	&SPB	&1.18	& 357	 &   5 &		\\
7630417	 & 123	&SPB	&2.03	& 613	 &  15 &		\\
8054179	 & 345	&SPB	&0.24:	& 114	 &  239&Fg, large $\mu$, high $g$\\
8087269$^\dagger$	 & 0123	&SPB	&1.61	& 1011   &  11 &Fg, possible Hybrid\\
8167938	 & 23	&SPB	&0.83	& 5725	 &   8 &		\\
8362546	 & 123	&SPB	&0.9	& 1287   &   2 &High $g$	\\
8766405$^\dagger$	 & 0123	&SPB	&1.83	& 1424   &   5 &Fg		\\
9075895	 & 2345	&SPB	&1.94	& 190	 &  12 &Fg		\\
9211123	 & 35	&SPB	&0.17	& 406	 &  39 &Fg		\\
9227988	 & 0123	&SPB	&0.17	& 2297   &   8 &		\\
9278405	 & 0123	&SPB	&1.78	& 9	 &  16 &Low amplitude	\\
9468611	 & 0123	&SPB	&2.14	& 33	 &   8 & Single Fg	\\
9582169	 & 0123	&SPB	&2.5	& 24	 &   2 &Low freq. noise	\\
9663239	 & 2345	&SPB	&0.38   & 280    &   2 & Possible sdB, large $\mu$\\
9715163	 & 0123	&SPB	&0.38	&   9	 &  29 &		\\
9715425	 & 0123	&SPB	&1.71	& 2628	 &  15 &Fg, low freq. noise\\
9884329	 & 0123	&SPB	&2.71	& 34	 &   5 &		\\
9910544	 & 0123	&SPB	&0.72	& 19	 &  23 &Fg, low freq. noise\\
10207025 & 35	&SPB	&1.45	& 54	 & 254 &Fg, noisy	\\
10220209 & 2	&SPB	&1.24	& 1299   &   9 &Possible Hybrid, large $g$\\
10526294 & 123	&SPB	&0.57	& 11133  &  10 &		\\
10658302$^\dagger$ & 0123	&SPB	&0.19	& 2127   &  12 &Fg, high $g$	\\
10790075 & 0123	&SPB	&0.26	& 51	 &  21 &Low freq. noise	\\
10797526$^\dagger$ & 0123	&SPB	&0.37	& 1842   &  19 &		\\
11152422 & 2345	&SPB:	&0.94	& 500	 &   5 &Possible A star	\\
11454304$^\dagger$ & 0123	&SPB	&0.62	& 397	 &  37 &Large $g$	\\
11671923 & 0123	&SPB	&1.72	& 38	 &  12 &		\\
11912716 & 123	&SPB	&1.51	& 97	 &   8 &		\\
11957098 & 0123	&SPB	&0.58	& 129	 &   8 &		\\
12021724 & 45	&SPB	&1.48	& 397	 &   3 & Modest $\mu$	\\
\\
1868650  & 0123	& BIN/ROT & 3.42  & 22946&   1 & Modest $\mu$	\\
2708156  & 0123	& BIN/ROT & 0.53  &110297&   8 &		\\
2853320  & 123	& BIN/ROT & 0.2	  & 3684 &   9 &		\\
2987640  & 123	& BIN/ROT & 0.23  &  343 &   2 &		\\
3216449  & 0123	& BIN/ROT & 1.06  & 2033 &   2 &		\\
3443582  & 45	& BIN/ROT & 0.92  & 3838 &   2 &Large $g$\\
3763002  & 123	& BIN/ROT & 0.58  &  245 &   32	&Possible SPB, low freq. noise\\
4064365  & 123	& BIN/ROT & 0.11  &15290 &   1 &		\\
4373805  & 0123	& BIN/ROT & 0.13  &  166 &  19 &		\\
4476114  & 12	& BIN/ROT & 0.34  & 2601 &   3 &		\\
4828345  & 0123	& BIN/ROT & 1.19  &14072 &   2 &High $g$	\\
4831616  &   23 & BIN/ROT & 0.48  & 4252 &   2 & \\
5352328  &   23 & BIN/ROT & 0.97  &  105 &   7:& \\ 
5450881  & 0123	& BIN/ROT & 3.68  &  934 &   4 &		\\
5473826  & 0123	& BIN/ROT & 0.95  &10800 &   2 &		\\
5479821$^\dagger$  & 0123	& BIN/ROT & 0.59  &12844 &   2 &		\\
5530935  & 0123	& BIN/ROT & 1.27  &   65 &   2 &		\\
5687841  & 0123 & BIN/ROT & 0.61  &   18 &  10 & Fg, possible SPB\\
5706079  & 123	& BIN/ROT & 0.24  & 3257 &   2	&		\\
5775128  & 0123	& BIN/ROT & 0.52  & 7677 &   4	&High $g$	\\
5942605  & 123	& BIN/ROT & 1.56  & 6593 &   2	&High $g$	\\
6278403  & 0123	& BIN/ROT & 0.84  & 2643 &   2	&		\\
6363494  & 23	& BIN/ROT & 1.1   &  872 &   4	&Possible A star\\
6864569  & 0123	& BIN/ROT & 0.43  &10933 &   2	&		\\
6878288  &   35 & BIN/ROT & 0.33  & 3552 &   5  & \\
6950556  & 12	& BIN/ROT & 1.32  & 6695 &   2	&		\\
6953047  & 23   & BIN/ROT & 0.58  & 5595 &   2 & \\
6954726$^\dagger$  & 0123	& BIN/ROT & 1.03  & 9076 &   43	&Low freq. noise   \\
7022658  & 23	& BIN/ROT & 0.64  & 5670 &   2	&Possible A star\\
7282265  & 23	& BIN/ROT & 0.89  & 3418 &   5	&		\\
7335517  & 35	& BIN/ROT & 7.29  &18094 &   4	&Large $\mu$	\\
7434250  & 25	& BIN/ROT & 0.24 &   549 &   65 &sdB?, low freq. noise, modest $\mu$\\
7599132$^\dagger$  & 0123	& BIN/ROT & 0.77  & 8056 &   2	&		\\
7749504  & 0123	& BIN/ROT & 3.53  & 1260 &   2	&		\\
7778838  & 0123	& BIN/ROT & 0.17  & 9732 &   1	&		\\
8129619  & 123	& BIN/ROT & 1.05  &  515 &   1	&		\\
8161798$^\dagger$  & 0123	& BIN/ROT & 0.91  &36288 &   2	&		\\
8183197  & 0123	& BIN/ROT & 0.27  &  264 &   1	&Low freq. noise	\\
8233804  & 123	& BIN/ROT & 1.51  & 5977 &   3	&High $g$	\\
8488717$^\dagger$  & 0123	& BIN/ROT & 0.31  &  247 &   1	&		\\
8619436  & 123	& BIN/ROT & 0.32  & 8884 &   1	&		\\
8754603  & 35	& BIN/ROT & 6.76  &11630 &   3	&Modest $\mu$	\\
8759258  & 0123	& BIN/ROT & 1.05  & 6509 &   3	&Modest $\mu$	\\
8871494  & 123	& BIN/ROT & 1.17  &  316 &   1	&		\\
9202990  & 34	& BIN/ROT & 0.08  &51338&   26	&Low freq. noise, large $\mu$, possible A star\\
9472174  & 0123	& BIN/ROT & 7.95  &53559 &   1	& 		\\
9958053  & 0123	& BIN/ROT & 0.73  & 3581 &   2	&		\\
10090722 & 0123	& BIN/ROT & 0.17  & 3033 &   4	&		\\
10130954$^\dagger$ & 0123	& BIN/ROT & 0.45  &  370 &   2	&High $g$	\\
10281890 & 0123	& BIN/ROT & 0.08  &13236 &   1	&		\\
10551350 & 2345	& BIN/ROT & 0.9   & 8095 &   2	&Large $\mu$	\\
10657664 & 0123	& BIN/ROT & 0.61  &  392 &  10	&Possible Hybrid	\\
10937527 & 123	& BIN/ROT & 1.31  & 7396 &   4	&Modest $\mu$, high $g$\\
10966623 & 345	& BIN/ROT & 1.38  & 4090 &   5	&High $g$\\\
11038518 & 123	& BIN/ROT & 1.31  & 3600 &   4	&		\\
11287855 & 2345	& BIN/ROT & 2.28  &  296 &   8	&Large $\mu$	\\
12207099$^\dagger$ & 0123	& BIN/ROT & 0.70  &14896 &   2	&		\\
12217324$^\dagger$ & 0123	& BIN/ROT & 0.15  &   17 &   22	&Very bright, low freq. noise\\
12268319 & 0123	& BIN/ROT  & 0.28  &  393 &   7	&Low freq. rise	\\
\\
2860851	&  0123	& H	&  3.71	& 346   & 10   &		\\
4547333	&   35	& H	&   		&          &        & Noisy spectrum, possibly nonvariable\\
5172153	&  23	         & H	&  3.14	& 621   & 20   &Possible $\delta$ Scuti, Fg\\
5643103	&  0123	& H	&  5.53	& 1100 &  4     &		\\
6777127	&  23	         & H	& 11.73	& 180   &  4     &Large $\mu$, possible A star\\
7108883	&  0123	& H	&  6.18	& 259   &  4     &Large $\mu$	\\
7664467 &   25            & H   &  0.25       & 163   &  60:   &Possible BCEP  \\
8264293	& 0123	& H	&  2.96	& 104    & 13    &		\\
10285114$^\dagger$ & 0123	& H	&  5.41	& 1282  &  7     &Fg, high $g$\\
11293898& 0123	& H	&  2.35	& 1486  & 18    &Fg		\\
11360704$^\dagger$ & 0123	& H	&  2.07	& 1972  & 16    &Fg		\\
11971405 & 0123	& H	&  4.01	& 3292   & 6     &Fg		\\
12116239 & 2345        & H   & 23.71      & 172    &59     &Likely A star \\
12258330$^\dagger$ & 0123	& H	&  2.49	& 662     &  4    &Large $\mu$	\\
\\
3527751	 & 25  	&sdB	&22.09	& 140     &   69 & Possible $\beta$~Cep star		\\
3561700	 & 23 	&sdB	& 1.05	& 3555   &   8   & Possible SPB, but large $\mu$\\
5807616	 & 25	         &sdB	&14.5	& 418     &   27 & Large $\mu$	\\
6188286	 & 2	        &sdB	&0.21	& 194     &   8   & Possible SPB	\\
6848529	 & 0123     &sdB	&	         & 	     & 	       &Large $\mu$, low freq. rise, sdB?\\
7098125	 & 23         &sdB	&0.34	& 129     &    6   &Large $\mu$, noisy	\\
7353409  &    25       &sdB & & & & Low freq. rise, modest $\mu$\\
7755741	 & 123       &sdB  & 0.34	& 229    &  17    & Possible H	\\
9543660	 & 123	&sdB	&	         & 	     &          &Large $\mu$, low freq. noise, high $g$, sdB?\\
10139564   & 25	         &sdB	&21.63	& 691     &  128	&Large $\mu$, low freq. rise\\
10904353   & 2345	&sdB	&1.45	& 28	     &    6:    &Large $\mu$	\\
10982905   & 34    	&sdB	&0.96	& 144     &   18	&Possible A star, noisy\\
\\
1718290	& 123	&WD	&  7.94	& 295	&  46  	&		\\
2697388	& 25	         &WD	&  17.33     & 1271	&  57:	&Low $g$, $\beta$~Cep?\\
3544662	& 12	         &WD	&  0.19:      &  94	         &  15:	&Low amp. signals, constant?\\
6866662	& 0123	&WD	&  3.22	& 626	&  9		&	Fg	\\
7540755	& 23    	&WD	&  3.80	& 482	& 13		&	Fg	\\
8619526	& 15	         &WD	&  0.89	& 142	&  6	&Large $\mu$, single freq?, low $g$\\
9020774	& 123	&WD	&  1.90	& 7796	&  4	& SPB?\\
10536147$^\dagger$ & 0123	&WD	&  0.22	& 1872	&  4	& SPB?\\
10799291& 123	         &WD	&  2.90	& 1908	& 13	& SPB?, Fg\\
11302565& 123	         &WD	&  1.37	& 539	& 15	& SPB?, Fg\\ 
11953267& 0123	&WD	&  2.70	& 2557       & 54	&Low freq. rise in power spectrum, poss Fg\\
\enddata
\tablenotetext{\dagger}{Denotes stars in common with \citet{balona2010}.}
\end{deluxetable}

\end{document}